\documentclass[twocolumn,superscriptaddress,secnumarabic,amssymb,%
amsmath,nobibnotes,aps,prd,showpacs,nofootinbib]{revtex4}
\usepackage{caption}
\usepackage{subcaption}
\usepackage{graphicx}
\usepackage{bm}
\usepackage{amsmath}
\usepackage{amsfonts}
\usepackage{amssymb}
\usepackage{epstopdf}
\usepackage{color}%
\providecommand{\U}[1]{\protect\rule{.1in}{.1in}}
\newcommand{\be}{\begin{equation}}
\newcommand{\ee}{\end{equation}}

\newcommand{\mincir}{\raise
-3.truept\hbox{\rlap{\hbox{$\sim$}}\raise4.truept\hbox{$<$}\ }}
\newcommand{\magcir}{\raise
-3.truept\hbox{\rlap{\hbox{$\sim$}}\raise4.truept\hbox{$>$}\ }}

\newtheorem{remark}{Remark}[section]

\begin{document}

\title{The Dapor-Liegener model of  Loop Quantum Cosmology: A dynamical analysis}

\author{Jaume de Haro\footnote{E-mail: jaime.haro@upc.edu}}
\affiliation{Departament de Matem\`atiques, Universitat Polit\`ecnica de Catalunya, Diagonal 647, 08028 Barcelona, Spain}

\thispagestyle{empty}

\begin{abstract}
The  Dapor-Liegener model of  Loop Quantum Cosmology (LQC), which depicts an emergent universe from a de Sitter regime in the contracting phase is studied from the 
mathematical viewpoint of dynamical systems and compared with the
standard model of LQC.  Dealing with perturbations, on the contrary to standard LQC where at early times all the scales are inside the Hubble radius, we show that it is impossible to implement the matter bounce scenario due to the fact that  an emergent de Sitter regime in the contracting phase implies that all  the scales are outside of the Hubble radius in a past epoch. 

\end{abstract}

\vspace{0.5cm}


\maketitle

\section{ Introduction}

In  a  recent paper \cite{dl} (see also \cite{dl2}),  applying Thiemann's procedure for the regularization of the 
 the full Hamiltonian in Loop Quantum Gravity (LQG), 
Dapor and Liegener  
have obtained a new effective Hamiltonian for Loop Quantum Cosmology (LQC) which agrees, at the leading order,  with the previous one obtained some years ago in
\cite{ma}, but differs from the usual effective Hamiltonian of LQC \cite{singh, singh1, singh2, singh08, singh09, as}. The difference between both approaches lies in the fact that
 for an spatially flat and homogeneous universe, the Euclidean and the Lorentz terms of the full Hamiltonian are proportional to each other and in LQC it is usual to write 
the Lorentz term as the Euclidean one and
 quantize their combination \cite{as}. However, this treatment is impossible in the
full LQG theory, where the Lorentz term has to be quantized in
a different way from of the  Euclidean one  \cite{dl1}, obtaining a completely different effective Hamiltonian.

\

This new effective Hamiltonian constraint leads, contrarily to standard LQC,  to a non-symmetric bouncing background emerging from a de Sitter regime in the contracting phase and ending in the expanding one by matching with General Relativity (GR) \cite{dl2}. And, 
although  this model has already been studied in great detail in several  papers \cite{dl, singh18, agullo}, 
we believe that an analysis from the viewpoint of dynamical systems   could simplify the reasoning and  help to better understand it.

\

In fact, working in the plane $(H,\rho)$ where $H$ denotes the Hubble parameter and $\rho$ the energy density of the universe, where the standard model in LQC has the universe crossing and ellipse in  clockwise direction, we show that the Dapor-Liegener model has a more complicated behavior
  presenting two separate 
asymmetric branches: In the physical  one, and always dealing with a non phantom field or fluid filling the universe, the universe emerges  from a de Sitter regime evolving, in the contracting phase, to the bounce, where after entering in the expanding phase it evolves asymtotically  into a flat  expanding universe obeying GR. In the non-physical one one has, at very early times,  a flat contracting universe obeying GR and evolving to the bounce, to enter in  the expanding phase where it transits  to end up in 
a de Sitter regime. This second branch is not physical, in spite of the  low value of the energy density of the universe, due
to the high value of the Hubble parameter in this last stage,  which  is in disagreement with its   very  low current value.

\

Once we have studied the dynamics of the model 
we deal with perturbations, arguing that due to the fact that  the universe emerges from  a  de Sitter regime in the contracting phase, and thus at early times
all the  scales are outside the Hubble radius,   it is impossible 
for this  new approach of LQC
to provide 
either a matter or matter-ekpyrotic bouncing scenario as the ones given by standard LQC, where at the beginning all the scales are inside of the Hubble radius \cite{we,ha14a, cw, wilson0}.

\

The paper is organized as follows: In Section II we review the dynamics of the standard LQC background. Section III is devoted to the analysis of the new approach of LQC
obtaining the corresponding modified Friedmann equation and the dynamical equations. Finally, in the last Section we briefly discuss some features of cosmological perturbations in this new scenario
using the so-called {\it dressed effective metric} approach \cite{agullo1}.

\

The units used throughout the paper are $\hbar=c=1$, and ${8\pi G}=1$.

\section{Dynamics in standard LQC}

We start reviewing the dynamics in standard LQC where 
the full effective Hamiltonian is given by \cite{as, singh, singh1, singh2, singh08, singh09}
\begin{eqnarray}\label{0}
{\mathcal H}_{LQC}={\mathcal V}\rho-3{\mathcal V}\frac{\sin^2(\beta\lambda)}{\gamma^2\lambda^2},
\end{eqnarray}
where 
$\rho$ is the energy density of the universe, $\gamma\cong 0.2375$ is the Immirzi parameter
whose numerical value
is obtained comparing the Bekenstein-Hawking formula with the black hole
entropy  calculated  in    LQG  \cite{meissner}, 
{{}
although an updated derivation  \cite{perez} shows
that the Immirzi parameter is no longer  fixed, but only bounded in the LQC setting, by this
formula.
The parameter}
$\lambda\equiv \sqrt{\frac{\sqrt{3}}{2}\gamma}$
is the square root of the {\it area gap} -the square root of the minimum eigenvalue of the area operator- in LQG 
{{} (see section II E of 	\cite{as} where the authors use an heuristic correspondence between the kinematic states of LQC and those of LQG to conclude that the parameter $\lambda$ is the square root of the minimum eigenvalue of the area operator of LQG), although
there are some modified theories leading to the same Friedmann and Raychauduri equations as in standard LQC, where $\lambda$ is a free parameter
which has to be determined from observational data. For example, 
teleparallel LQC   \cite{saridakis, haro13}, theories including 
  in the Einstein-Hilbert action
 a convenient non-linear  term of the form $f({\mathcal R})$,  where ${\mathcal R}$ is some scalar such that in the Friedmann-Lema{\^\i}tre-Robertson-Walker  (FLRW) spacetime  becomes proportional to the Hubble parameter or its square
 \cite{helling, ds09, ha17}   
  or else using a modified version of mimetic gravity \cite{hap,Norbert,mukhanov1,langlois}. } 
Finally, ${\mathcal V}=a^3$ is the volume (to simplify the volume of the cubic fiducial cell has taken to be equal to $1$)  and $\beta$ is its conjugate momentum, which classically satisfies $\beta=\gamma H$ \cite{acs}, being $H$  the Hubble parameter, although 
and whose Poisson bracket is given by  $\{\beta,{\mathcal V}\}=\frac{\gamma}{2}$.

\

The Hamiltonian constraint $ {\mathcal H}_{LQC} =0$, leads to the following expression of the energy density
\begin{eqnarray}\label{01}
\rho=3\frac{\sin^2(\beta\lambda)}{\gamma^2\lambda^2},
\end{eqnarray}
and 
the Hamilton equation $\dot{{\mathcal V}}=\{{\mathcal V}, {\mathcal H}_{LQC}\}=-\frac{\gamma}{2}\frac{\partial {\mathcal H}_{LQC}}{\partial \beta}$
leads to the following expression for the Hubble parameter
\begin{eqnarray}\label{02}
H=\frac{\sin(2\beta\lambda)}{2\gamma \lambda}.
\end{eqnarray}

A simple combination of equations  (\ref{01}) and (\ref{02}) leads to the  holonomy corrected Friedmann equation in standard LQC \cite{singh06, svv06,sst06}
\begin{eqnarray}\label{03}
H^2=\frac{\rho}{3}\left(1-\frac{\rho}{\rho_c}\right),
\end{eqnarray}
where $\rho_c=\frac{3}{\gamma^2\lambda^2}\cong 252$  is the so-called {\it critical  {{} energy} density} in standard LQC \cite{singh08}.  From this 
modified Friedmann equation one can see that at low energy densities ($\rho\ll \rho_c$) one recovers GR, because this equation becomes the standard
Friedmann equation  $H^2=\frac{\rho}{3}$ which depicts a parabola in the plane $(H, \rho)$. 
This two curves which at low energy densities coincide,  are very different at high energy densities. Effectively, the parabola of GR is unbounded allowing the formation of 
singularities such as the Big Bang or the Big Rip where the energy density diverges. However, in standard LQC,  this kind of singularities are forbidden due to
the fact that the ellipse depicted by the Friedmann equation in standard LQC (see FIG. $1$)  is a closed bounded curve \cite{bho, aho}.

\

To find the dynamical equation,  we have to take into account that  holonomy corrections only affect the gravitational sector, for this reason 
the energy density satisfy  the conservation equation
$\dot{\rho}=-3H(\rho+P)$, where $P$ is the pressure. Then, taking the derivative of (\ref{03}) and using the conservation equation
one can easily find the Raychauduri equation in standard LQC \cite{singh09}
\begin{eqnarray}
\dot{H}=-\frac{1}{2}(\rho+P)\left(1-\frac{2\rho}{\rho_c}  \right).
\end{eqnarray}

Note that from the conservation equation one can see that for a fluid or field with effective Equation of State (EoS) parameter $w_{eff}=\frac{P}{\rho}>-1$, that is, for a non-phantom
fluid or field, the movement accros the ellipse is clockwise, as has been shown in FIG. $1$.

\begin{figure}[h]
\begin{center}
\includegraphics[scale=0.5]{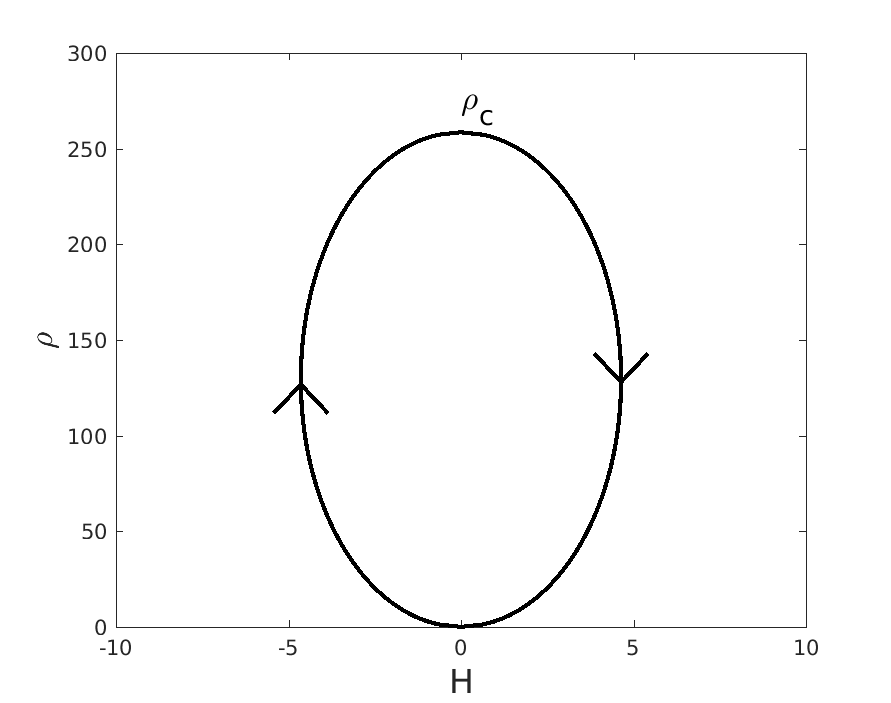}
\end{center}
\caption{Ellipse depicted by the Friedmann equation in standard LQC, and its dynamics for a non-phantom  fluid or field.}
\end{figure}

\

Once we have obtained the dynamical equations, we can consider two different cases:
\begin{enumerate}
\item A universe filled by a barotropic fluid with EoS $P=P(\rho)$. In this case, the unique background is obtained solving the
first order differential equation
$\dot{\rho}=-3H_{\pm}(\rho)[\rho+P(\rho)],$
where $H_+(\rho)=\sqrt{\frac{\rho}{3}\left(1-\frac{\rho}{\rho_c}\right)}$ is the value of the Hubble parameter in the expanding phase 
and $H_-(\rho)=-\sqrt{\frac{\rho}{3}\left(1-\frac{\rho}{\rho_c}\right)}$ is its value in the contracting one.
In general, this equation has to be solved numerically, but in the particular case of an constant effective EoS parameter 
$w_{eff}$ one obtains the following analytic solution \cite{we}:
\begin{eqnarray}\label{lqcbackground}
a=\left(\frac{3}{4}\rho_c(1+w_{eff})^2t^2+1  \right)^{\frac{1}{3(1+w_{eff})}}, \nonumber\\ \quad H=\frac{\rho_c (1+w_{eff}) t} {2  a^{3(1+w_{eff})} }, \quad 
\rho=\frac{\rho_c}{a^{3(1+w_{eff})}}.
\end{eqnarray}

\item A universe filled by an scalar field  $\phi$ minimally coupled with gravity. In this case the energy density is $\rho=\frac{\dot{\phi}^2}{2}+V(\phi)$,
and the conservation equation reads
\begin{eqnarray}\label{conservation}
\ddot{\phi}+3H_{\pm}(\phi,\dot\phi)\dot{\phi}+V_{\phi}=0,
\end{eqnarray}
where once again $H_{\pm}(\phi,\dot\phi)=\pm\sqrt{\frac{\rho}{3}\left(1-\frac{\rho}{\rho_c}\right)}$.

The difference with the case of a barotropic fluid is that now we have a second order differential equation, meaning
that one has infinitely many backgrounds. Moreover, one could also obtain a potential having a background which is
the same as the one provided by a barotropic fluid with constant EoS parameter \cite{mielczarek, haa}
\begin{eqnarray}\label{potlqc}
V=2\rho_c\frac{(1-w_{eff})e^{-\sqrt{3  (1+w_{eff}) }\phi}}{(1+e^{-\sqrt{3(1+w_{eff})}\phi})^2}.
\end{eqnarray}
\end{enumerate}

Effectively, inserting this potential in (\ref{conservation}) one gets the analytic solution
\begin{eqnarray}
{\phi}=\frac{2}{\sqrt{3(1+w_{eff})}}\ln\left( \sqrt{\frac{3}{4}\rho_c  (1+w_{eff})^2 }t   \nonumber\right. \\ \left.+\sqrt{\frac{3}{4}\rho_c   (1+w_{eff})^2   t^2+1}  \right),
\end{eqnarray}
which leads to the background (\ref{lqcbackground}).  The dynamics provided by the potential (\ref{potlqc}), i.e., the other non-analytic solutions, was recently studied with great detail in \cite{aah, haa}, showing that in the case $|w_{eff}|<1$ all the backgrounds depict a universe with a constant effective EoS parameter equal to $w_{eff}$  at early and late times. On the contrary, when $w_{eff}>1$, the potential (\ref{potlqc}) becomes ekpyrotic, the backgrounds depict an universe bouncing twice and after the second bounce it  enters, in the expanding phase, in a kination regime (its effective EoS parameter is equal to $1$).  

\

\section{Dynamics in the Dapor-Liegener model of LQC}

In the   Dapor-Liegener (DL) model the full effective Hamiltonian  is given by \cite{dl, ma, dl2, singh18, agullo}
\begin{eqnarray}\label{1}
{\mathcal H}_{DL}={\mathcal V}\rho-3{\mathcal V}\frac{\sin^2(\beta\lambda)}{\gamma^2\lambda^2}\left(1-(\gamma^2+1)\sin^2(\beta\lambda)\right),
\end{eqnarray}
and  the Hamiltonian constraint leads to the following expression of the energy density of the universe
\begin{eqnarray}\label{4}
\rho=3\frac{\sin^2(\beta\lambda)}{\gamma^2\lambda^2}\left(1-(\gamma^2+1)\sin^2(\beta\lambda)\right).
\end{eqnarray}

In this model, the Hamilton equation $\dot{{\mathcal V}}=\{{\mathcal V}, {\mathcal H}_{DL}\}=-\frac{\gamma}{2}\frac{\partial {\mathcal H}_{DL}}{\partial \beta}$
leads to the following value of the Hubble parameter
\begin{eqnarray}\label{3}
H=\frac{\sin(2\beta\lambda)}{2\gamma \lambda}\left(1- 2(\gamma^2+1)\sin^2(\beta\lambda)\right).
\end{eqnarray}

Introducing the notation $\sin^2(\beta\lambda)\equiv x$, 
the equation (\ref{3}) has the form ${\rho}=\rho_c f(x)$,
where $f$ is a function defined in $[0,1]$ as $f(x)= x-(\gamma^2+1)x^2$. This function is  positive in the interval $[0, \frac{1}{\gamma^2+1} ]$, and reaches its
maximum at $x=\frac{1}{2(\gamma^2+1)}$, meaning that the minimum value of the energy density is $0$ and its maximum value is $\rho_{max}= \frac{\rho_c}{4(\gamma^2+1)}$.

\

Using this variable $x\in [0, \frac{1}{\gamma^2+1}]$ the equation (\ref{3}) could be written as
\begin{eqnarray}\label{5}
H^2=\frac{x(1-x)\rho_c }{3}\left(1-2(\gamma^2+1)x\right)^2,
\end{eqnarray}
which in the interval  $[ 0, \frac{1}{\gamma^2+1}]$  vanishes when $x=0$, $x=1$ and $x=\frac{1}{2(\gamma^2+1)} $ i.e., when $\rho=0$ and $\rho=\rho_{max}$. Note also that, when the energy density vanishes at $x=\frac{1}{\gamma^2+1}$, the square of the Hubble parameter does not vanish as in standard LQC. In this theory its value is 
$\tilde{H}^2=
\frac{4\gamma^2\rho_{max}}{3{\gamma^2+1}}
$.

\

After this brief discussion one can conclude that
 the variable $\beta$ belongs in the interval $[-\beta_i,\beta_i]$ where $\beta_{i}\equiv \frac{1}{\lambda}\arcsin\left(\frac{1}{\sqrt{\gamma^2+1}} \right)$. 
The equations (\ref{3}) and (\ref{4}) depict a curve in the plane $(H,\rho)$ whose 
  first branch  is obtained when $\beta$ belongs in the interval $[0,\beta_i]$ and the second one when the variable belongs in $[-\beta_i,0]$ (see FIG. $2$).

\

To find the dynamics we perform the temporal derivative of the energy density (\ref{4}) obtaining:
\begin{eqnarray}\label{6}
\dot{\rho}=
\frac{6}{\gamma}H\dot{\beta},
\end{eqnarray}
and once again  taking into account that holonomy corrections only affect the matter sector,  the conservation equation  will be $\dot{\rho}=-3H(\rho+P)$,  and one 
 finally gets the equation
\begin{eqnarray}\label{7}
\dot{\beta}=
-\frac{\gamma}{2}(1+w_{eff})\rho,
\end{eqnarray}
where once again $w_{eff}=\frac{P}{\rho}$ denotes the effective EoS parameter.

\

Since the derivative of $\beta$ is zero when the energy density vanishes, the dynamical system has three fixed points at $\beta=\pm \beta_i$ and $\beta=0$, or in the plane $(H,\rho)$, at $(\pm\sqrt{\tilde{H}^2},0)$ and $(0,0)$. 
Therefore, for a non-phantom fluid or field, i.e., when $w_{eff}>-1$, 
there are two different dynamics:
\begin{enumerate}
\item Branch 1 (blue curve in FIG. $2$):
The variable $\beta$ moves from $\beta_i$ to $0$, or in the plane $(H,\rho)$ the universe emerges in a de Sitter regime   moving  in the contracting phase  
from $(-\sqrt{\tilde{H}^2},0)$  to $(0,\rho_{max})$,
where it bounces to enter in the expanding phase, and finally at late times it ends at  $(0,0)$ where GR applies, because 
 when $\beta\cong 0$ the equations (\ref{3}) and (\ref{4}) become $H\cong \frac{\beta}{\gamma}$ and $\rho\cong 3\frac{\beta^2}{\gamma^2}$, thus combining them one gets the standard Friedmann equation $H^2\cong \frac{\rho}{3}$, that is, one recovers  GR at low energy densities.

\item  Branch 2  (red curve in FIG. $2$):  The variable $\beta$ moves from $0$ to $-\beta_i$, or in the plane $(H,\rho)$ the universe starts when GR is valid, moving in the contracting phase from
$(0,0)$  to $(0,\rho_{max})$,  where the universe bounces entering in the expanding phase, and finally 
ending  in a de Sitter regime at $(+\sqrt{\tilde{H}^2}, 0)$. Obviously, the dynamics in this second branch is not viable because the universe  ends in a de Sitter phase with such a large value of the Hubble parameter.
\end{enumerate}

\

\begin{figure}[h]
\begin{center}
\includegraphics[scale=0.5]{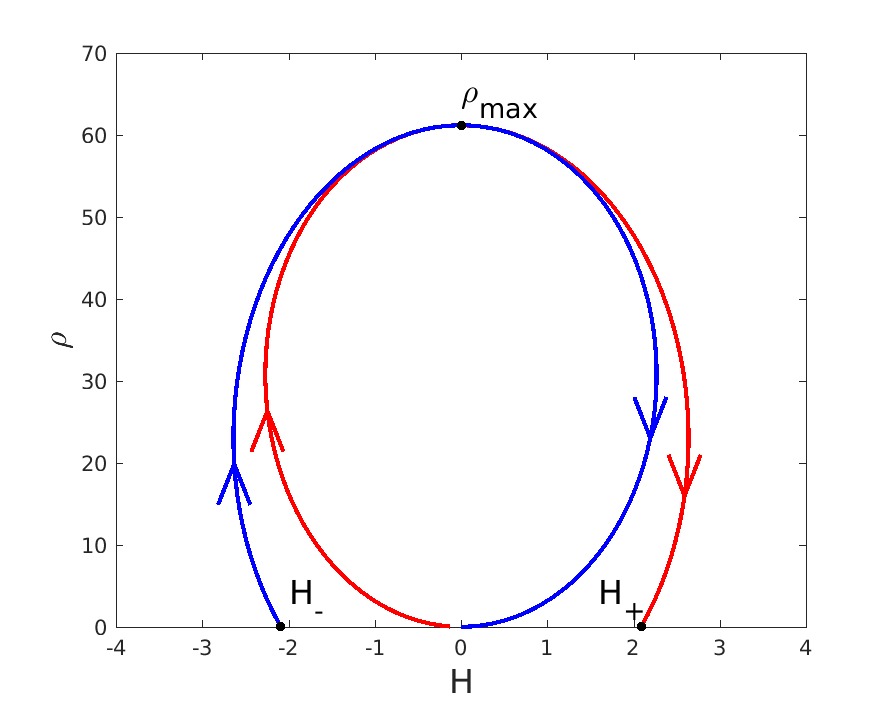}
\end{center}
\caption{Curve depicted by the Friedmann equation in the Dapor-Liegener model of LQC,
and its dynamics for either a non-phantom fluid or field.
Blue: 
branch with $\beta>0$. Red: branch with $\beta<0$.}
\end{figure}

\

To obtain explicitly the dynamics one has to solve the equation (\ref{7}). In the case of a barotropic fluid with EoS $P=P(\rho)$ one has to solve the equation
\begin{eqnarray}\label{beta1}
\dot{\beta}=-\frac{\gamma}{2}(\rho+P(\rho)),
\end{eqnarray}
with $\rho$ given by equation (\ref{4}). This is a one dimensional first order differential equation in the variable $\beta$ which, once it is solved, one has to
insert in equations (\ref{4}) and (\ref{3}) to obtain the dynamics. 
In the particular case of a constant effective EoS parameter, the equation (\ref{beta1}) can be integrated analytically obtaining an implicit equation of the form
$F(\sin(\beta\lambda))=t$ but, unfortunately, there is no analytic expression of the inverse of $F$. Therefore, it is impossible to reach  a simple expression such as 
(\ref{lqcbackground}) obtained in standard LQC, only numerical calculations can be performed to obtain the dynamics.

\

When the universe is filled by an scalar field  minimally coupled with gravity the problem is more involved because in this case, equation (\ref{7}) reads
\begin{eqnarray}
\dot{\beta}=-\frac{\gamma}{2}\dot{\phi}^2,
\end{eqnarray}
and it is impossible to express $\dot{\phi}^2$ as a function of $\beta$. So, one has to work as in standard LQC, and consider the conservation equation
(\ref{conservation}), but with another expression of $H_{\pm}(\phi,\dot{\phi})$. To find it, one has to isolate $\sin^2(\beta\lambda)$ in (\ref{4}) and insert it in (\ref{3}).
After some algebra one has 
\begin{eqnarray}\label{friedmannlqc}
H^2_{\pm}(\rho)=\frac{\rho}{3(\gamma^2+1)}\left(1-\frac{\rho}{\rho_{max}}  \right)
\times \nonumber\\
\left[1+\frac{2\gamma^2}{1\pm \sqrt{1-\frac{\rho}{\rho_{max}}}}  \right],
\end{eqnarray}
and since at low energy densities $\rho\ll\rho_{max}$ one has $H^2_{+}(\rho)\cong \frac{\rho}{3}$ and $H^2_{-}(\rho)\cong \tilde{H}^2$, the dynamics in the first branch (the physical one) will be given by the equation
\begin{eqnarray}\label{conservation1}
\ddot{\phi}+3H_{\pm}(\phi,\dot\phi)\dot{\phi}+V_{\phi}=0,
\end{eqnarray}
where we now 
denote by $H_{+}(\phi,\dot\phi)$ the value of the Hubble parameter in the expanding phase and $H_{-}(\phi,\dot\phi)$ in the contracting one, and we have
$H_{\pm}(\phi,\dot\phi)=\pm \sqrt{H^2_{\pm}(\rho)}$, with $\rho=\frac{\dot{\phi}^2}{2}+V(\phi)$. One can see that equation (\ref{conservation1}), which as in GR or standard LQC  provides infinitely many different backgrounds because it is a second order differential equation,  can only 
be solved numerically.

\

Another way, the one used in \cite{agullo},  to find numerically backgrounds is to consider the system 
\begin{eqnarray}
\left\{\begin{array}{ccc}
\dot{\beta} & = &-\frac{\gamma}{2}\psi^2\\
\dot{\phi} &=& \psi\\
\dot{\psi}&=&-3H(\beta){\psi}-V_{\phi},
\end{array}
\right.
\end{eqnarray}
where $H(\beta)$ is given by (\ref{3}).  To solve the system one needs three initial conditions, which for simplicity one could take at the bounce,
 $(\beta_B, \phi_B, {\psi}_B)$. As we have already showed, in the first branch,  at the bounce one has $\beta_B=\frac{1}{\lambda}\arcsin\left(\frac{1}{\sqrt{2(\gamma^2+1)}}  \right)$, then one only has to choose a value of $\phi_B$ satisfying $V(\phi_B)\leq \rho_{max}$, because, at the bounce, $\psi_B$ is determined by the constraint
 $\frac{{\psi}_B^2}{2}+V(\phi_B)=\rho_{max}$.

\

Finally,  we want to stress that the background provided by the Dapor-Leigener model  does not seem easy to be mimicked using modified or mimetic gravity as has been done for the standard model in LQC \cite{helling, Norbert, langlois}, due to the complicated form exhibited by the solution curve to the Friedmann equation (\ref{friedmannlqc}) 
in the plane  $(H,\rho)$ (see FIG. $2$).

\section{Perturbations}

There are two different ways to understand a bouncing scenario (see \cite{B,nb08,patrick0,patrick,odintsov0} for a review of bouncing cosmologies). One of them is to see it as an implementation of inflation, where  the big bang singularity is replaced by a bounce but the inflationary regime exists in the expanding phase \cite{bgt, bgt1, barrau, sloan}. The other viewpoint is radically different: a bouncing cosmology,  {{}  named {\it matter or matter-ekpyrotic bouncing scenario}}, is an alternative to the inflationary paradigm, and thus, the inflationary phase is removed
\cite{brandenberger, brandenberger1,brandenberger2} in this scenario.

\

In the first path, an inflationary potential is used and the  observable scales leave the Hubble radius during the inflationary regime as in standard inflation but, unlike in inflation, the modes corresponding to those scales 
are not expected to be in the so-called {\it adiabatic} or, sometimes, Bunch-Davies vacuum (see for instance the section $6.2$ of \cite{riotto}), due to their previous
evolution in the contracting phase and across the bounce \cite{agullo}. On the contrary, in the second point of view the
modes corresponding to the observable scales, which leave the Hubble radius at very early times in the contracting phase,  are in the {\it adiabatic} vacuum \cite{ we, wilson0,cw,ha14a,saridakis,ha17, lw}, due to the duality between the de Sitter regime in the expanding phase and a matter domination in
the contracting phase \cite{wands}. In fact,  
to obtain a nearly flat power spectrum in this approach,  one has to choose a  potential which at early times leads to a quasi-matter domination
regime in the contracting era \cite{eho}. 

\

In standard LQC, both points of view have been implemented with success. The first one has been extensively studied in \cite{bgt, bgt1, agullo1, agullo2, agullo3}, and the second one in \cite{we, wilson0, cw, haa, eho, lw, ha14a, ha17} studying the matter and the matter-ekpyrotic bounce scenario, and partially showing its viability confronting the theoretical values of spectral quantities such as the spectral index, its running or the ratio of tensor to scalar perturbations with their corresponding observational values.

\

However, as we will immediately show, in this new version of LQC it is impossible to implement the second viewpoint. Effectively, 
{{}
in  LQC  there are two ways to deal with perturbations,  {\it the deformed algebra approach} 
\cite{grain, vidotto, lisenfors, grain1}
and {\it the dressed effective metric approach}
\cite{agullo1, agullo2, agullo3}. Both approaches are performed {  in the Hamiltonian framework instead of the Lagrangian one, so
 covariance is not immediately manifest and it is replaced  by the invariance under gauge transformations generated by the Hamiltonian constraint $H[N]$ and 
diffeomorphism constraints $D[N^a]$, where $N$ is the smeared lapse function and $N^a$ is the smeared shift vector, which satisfy the classical algebra of constraints:
\begin{eqnarray}\label{x1}
\{D[N_1^a], D[N_2^a]\}= 
D[{\mathcal L}_{N_1} N_2^a], 
\end{eqnarray}
\begin{eqnarray}\label{x2}
\{H[N], D[M^a]\}= 
-H[{\mathcal L}_{M} N],
\end{eqnarray}
\begin{eqnarray}\label{x3}
\{H[N_1],H[N_2]\}= D[ q^{ab}( N_1\nabla_a N_2-N_2 \nabla_a N_1) ].
\end{eqnarray}

Then,  general covariance in canonical theories is implemented in a  more subtle way than in Lagrangian theories. And dealing with the quantization of canonical theories of gravity, two assumption are imposed in order to maintain the covariance (see for instance \cite{bojowald0,bojowald1,bojowald2}):
\begin{enumerate} 
\item The algebra of constraints has to be closed
\item The algebra of constraints has to be a well defined classical limit, and in the limit it has to coincide with the classical one: equations  (\ref{x1}), (\ref{x2}) and (\ref{x3}).
\end{enumerate}

\

 In the {\it deformed algebra approach} applied to standard LQC,    the Asthekar connection is replaced by hand by suitable sinusoidal functions \cite{grain}, and the anomalies, which appear after the replacement, are removed 
 introducing some counter-terms. The   obtained   algebra of constrains differs  from the classical case in the constraint
 \begin{eqnarray}
 \{H[N_1],H[N_2]\}=\Omega D[ q^{ab}( N_1\nabla_a N_2-N_2 \nabla_a N_1) ],
 \end{eqnarray}
 where $\Omega=1-\frac{2\rho}{\rho_c}$. Thus, in the classical limit $\rho_c\rightarrow \infty \Longleftrightarrow \gamma\rightarrow 0$, one recovers the
 classical expression, what means that  the {\it deformed algebra approach}  satisfy both assumptions, and consequently this approach maintains covariance.
 
 On the contrary, in  the {\it dressed  effective metric approach} where  the Mukhanov-Sasaki is the same as in GR, but the metric background is replaced by an effective one which differs from the classical one. In fact, the metric background is replaced by the one provided by LQC, what does not seem to preserve the covariance. } 

\

On the other hand, there are covariant theories, performed from a Lagrangian formulation,  which leads to the same background as standard LQC \cite{hap, hae18}, but the perturbation equations are completely different of those of LQC, { this is a point that deserves future investigation because is not clear at all why this equations
differs from the ones of the {\it deformed algebra approach} which, as we have already seen, is also covariant.}

\

Dealing with the Dapor-Liegener model of LQC, due to the difficult form of the corresponding Friedmann equation (eq. (\ref{friedmannlqc})), so far  the perturbation equations are not obtained
either in  the {\it deformed algebra approach} or  in any covariant  { Lagrangian} formulation. For this reason, although it  not { seems} covariant, to deal with the perturbed equations in the DP model of LQC, at the present time,  one has to use the {\it dressed effective metric approach}. { However, as we will immediately see, 
the chosen perturbative approach will not affect our claim about the impossibility  to implement the matter bounce scenario in the DP model of LQC, because in this scenario the observable modes must leave the Hubble radius at very low energy density where holonomy corrections become  negligible, and thus, during this period,  the perturbative equations
become the same as in GR. }

\

Thus,} studying scalar perturbations in this last approach, 
the Mukhanov-Sasaki (M-S) equation will   be in conformal time \cite{agullo3}
\begin{eqnarray}\label{msequation}
v_k''+\left(k^2+\mathfrak{U}-\frac{a''}{a}\right)v_k=0,
\end{eqnarray}
where the potential $\mathfrak{U}$ is given by
\begin{eqnarray}
\mathfrak{U}=\left(V\frac{3\dot{\phi}^2}{\rho}-2V_{\phi}\sqrt{\frac{3\dot{\phi}^2}{\rho}}+V_{\phi\phi}\right)a^2.
\end{eqnarray}

\

Dealing, for instance,  with a quartic chaotic potential $V=\lambda \phi^4$, at very early times, i.e., when $\rho\cong 0$, and thus with 
$\phi\cong 0$ and $\dot{\phi}\cong 0$,  one will have $\mathfrak{U}\cong 12\lambda \phi^2 a^2$.

\ 

\begin{remark}
The backgrounds provided by power law potentials $V=\lambda \phi^{2n}$, which has been reproduced numerically for the particular case of a  quadratic potencial (see for instance \cite{singh18}),  are not difficult to understand in LQC. At very early times, since the energy density is zero,  the field is at the bottom of the potential starting to oscillate to gaing energy because it is in the contracting phase (recall the conservation equation $\dot{\rho}=-3H\dot{\phi}^2$). In fact, in the Dapor-Liegener model, contrary to standard LQC, due to the high value of the Hubble parameter in the de Sitter regime, it only needs few oscillations to leave the minimum of the potencial and start to climb up the potential to reach the 
maximum of energy density and enter the expanding phase, where it rolls down the potential to finish oscillating once again at the bottom of the potential.
\end{remark}
\

On the other hand, recalling that 
 we only consider the first branch because, as we have already discussed, is the only physically viable,  at early times the universe is in a de Sitter regime in the contracting phase, meaning that the scale factor evolves as $a(t)=\tilde{a}e^{tH_-}$, and clearly, $\lim_{t\rightarrow-\infty}a(t)=\infty$. The conformal time is given by
 \begin{eqnarray}
 \tau= \int\frac{1}{a(t)}dt\quad \Longrightarrow \quad \tau=\frac{-1}{aH_-},
 \end{eqnarray}
and thus, $\lim_{t\rightarrow -\infty}\tau=0$, which is completely different to what happens with a de Sitter regime in the expanding phase, because in this case
if one denotes by  $H_S>0$  the value of the Hubble parameter, one has $\tau=\frac{-1}{aH_S}$,
and thus,  $\lim_{t\rightarrow -\infty}\tau=-\infty$.

\

{ This difference, affects directly the M-S equation,
which  in the {\it dressed effective metric} approach or any other approach, at early times,   has the same approximate form as in GR}
\begin{eqnarray}\label{msequation1}
v_k''+\left(k^2-\frac{2}{\tau^2}\right)v_k=0,
\end{eqnarray}
because $\mathfrak{U}\sim \frac{\phi^2}{\tau^2}\ll \frac{2}{\tau^2}$ ($\phi\cong 0$ at very early times) and in standard inflation
the M-S equation is \cite{riotto}
\begin{eqnarray}
v_k''+\left(k^2-\frac{z''}{z}\right)v_k=0,
\end{eqnarray}
where $z=\frac{a\dot{\phi}}{H}=\sqrt{2\epsilon}a$, being $\epsilon$ the main slow roll parameter.

\begin{remark}
The same result is obtained for the quadratic potential $V=\frac{1}{2}m^2\phi^2$, because in this case, at very early times,  one has 
$\mathfrak{U}\cong m^2a^2 $ with $m^2\sim 10^{-11}$. Effectively, in inflation the power spectrum of scalar perturbations is given by \cite{basset}
\begin{eqnarray}
{\mathcal P}\cong \frac{H_*^2}{8\pi^2\epsilon_*}\sim 2\times 10^{-9},
\end{eqnarray}
where the star means that the quantities are evaluated when the pivot scale leaves the Hubble radius. 
Since for the quadratic potential  the slow roll parameters satisfy $\epsilon_*=\eta_*=\frac{2}{\phi^2_*}$, and the spectral index is given by $n_s-1=-6\epsilon_*+2\eta_*$ \cite{basset} one gets \begin{eqnarray}
m^2\sim 3\pi^2(1-n_s)^2\times 10^{-9}\cong 4\times 10^{-11},
\end{eqnarray}
were, as usual,  we have taken $n_s=0.96$ (see for instance \cite{planck, planck1}).
\end{remark}

\

Then, in the contracting phase, as we have already shown,  { in the DL model of LQC} the conformal time starts at $\tau=0$ and then increases, which means that at the beginning all the modes are outside of the Hubble
radius and they enter into it, which is the contrary to what happens in a de Sitter regime in the expanding phase, where at the beginning the conformal time is $-\infty$, and thus, the modes leave the Hubble radius. 

\

For this reason it is impossible to 
{{}  implement the matter o matter-ekpyrotic bouncing scenario
}
in the physical branch of the new LQC model, because it is needed that the observable scales leave the Hubble radius at very early times. Moreover, there is a more conceptual problem in order to define the vacuum modes. Effectively, the general solution of (\ref{msequation1}) is a combination of Hankel functions
\begin{eqnarray}
&& v_k(\tau) =-\sqrt{\frac{\pi\tau}{4}}\left(C_1(k)H_{{3}/{2}}^{(1)}(k\tau)+C_2(k)H_{{3}/{2}}^{(2)}(k\tau)\right)=\nonumber\\
&&C_1(k)\frac{e^{ik\tau}}{\sqrt{2k}}\left(1+\frac{i}{k\tau}\right)+
C_2(k)\frac{e^{-ik\tau}}{\sqrt{2k}}\left(1-\frac{i}{k\tau}\right).
\end{eqnarray}

\
Therefore, 
when the de Sitter regime is in the expanding phase, and all the modes are inside the Hubble radius, the general solution  of  (\ref{msequation1}) is approximately equal to
\begin{eqnarray}
v_k(\tau)=
C_1(k)\frac{e^{ik\tau}}{\sqrt{2k}}+
C_2(k)\frac{e^{-ik\tau}}{\sqrt{2k}},
\end{eqnarray}
and one can choose the vacuum mode taking $C_1(k)=0$ and $C_2(k)=1$ as in the Minkowskian spacetime, because   the modes  well inside the Hubble radius  {\it do not feel} gravity. On the contrary, when the de Sitter regime is in the contracting phase, at very early times, all the modes are outside the Hubble radius, and the approximate form of the general solution of  (\ref{msequation1}) is 
\begin{eqnarray}
v_k(\tau)=
\left(C_1(k)
-C_2(k)\right)\frac{1}{\sqrt{2k}}\frac{i}{k\tau},
\end{eqnarray}
and, from our viewpoint,  it not clear at all how to choose the coefficients $C_1(k)$ and $C_2(k)$. Of course,
the more natural choice seems the same as  in a de Sitter regime in the expanding phase, as has been argued in \cite{agullo}, but without the same justification as in inflation  because at very early times all modes are
outside the Hubble radius feeling gravity.

\

To end this Section, we will  calculate the range of values of the pivot scale $k_*$ in co-moving coordinates. The relation with its physical value at time $t$, namely 
$k_{phys}(t)$ is given by
$k_*=a(t)k_{phys}(t)$.  The physical value, at the present time,  used by the Planck's team is $k_{phys}(t_0)=10^2H_0$  \cite{planck1}, where the sub-index $0$ means present time. Then, denoting   the  beginning of the radiation era and the equilibrium matter-radiation by the sub-index $R$ and $eq$. We will have
\begin{eqnarray}
k_*=10^2H_0a_0=10^2H_0\frac{T_{eq}}{T_0}a_{eq},
\end{eqnarray}
where $T_0$ and $T_{eq}$ are the corresponding  CMB radiation temperatures,  and where we have also used that the evolution is adiabatic (the entropy is conserved) after equilibrium. We now use that during the radiation epoch one has 
$\left(\frac{a_R}{a_{eq}}  \right)^4=\frac{\rho_{eq}}{\rho_R}$, and the formulas \cite{rehagen}
 \begin{eqnarray}
 \rho_{eq}\cong \frac{\pi^2}{15}g_{eq}T_{eq}^4, \qquad  \rho_R\cong \frac{\pi^2}{30}g_RT_R^4,
 \end{eqnarray}
where $g_{eq}\cong 3.36$ and $g_R$ depends on the reheating temperature. Then, we have
\begin{eqnarray}
k_*=10^2\frac{H_0}{T_0}\left(\frac{g_R}{2g_{eq}} \right)^{1/4}T_R a_R.
\end{eqnarray}

Now, dealing with an inflationary power law potential $V=\lambda \phi^{2n}$, where the universe is reheated via particle production due to the  oscillations 
of the inflaton field \cite{linde}. After inflation, the universe evolves, up to reheating, in a regime with constant effective EoS parameter given by $w_{eff}\cong \frac{n-1}{n+1}$
\cite{turner, ford}. For the sake of simplicity, we consider a quadratic potencial, so after reheating the universe evolves as matter dominated universe. Then, denoting by $end$ the end of the slow-roll period one will have  $ \left(\frac{a_{end}}{a_R}  \right)^3=\frac{\rho_R}{\rho_{end}}$, and thus
\begin{eqnarray}
k_*= 10^2\frac{\left(\frac{15}{\pi^2}\right)^{1/3}}{(2g_{eq})^{1/4}(2g_R)^{1/12}}\frac{H_0}{T_0}\left(\frac{\rho_{end}}{T_R}\right)^{1/3}a_{end} \nonumber\\
\cong \frac{70}{(2g_R)^{1/12}}\frac{H_0}{T_0} \left(\frac{\rho_{end}}{T_R}\right)^{1/3}  a_{end}. 
\end{eqnarray}

Let $N_B$ be the number of e-folds from the bounce to the end of the slow-roll phase. Then,  taking the scale factor equal to $1$ at the bounce - we can do it because we are dealing with geometries with  spatially  flat sections- we obtain the formula 

\begin{eqnarray}
k_* 
= \frac{70}{(2g_R)^{1/12}}\frac{H_0}{T_0} \left(\frac{\rho_{end}}{T_R}\right)^{1/3}  e^{N_B}. 
\end{eqnarray}

In this formula, $\rho_{end}$ and $N_B$ are calculated from the background. Effectively, inflation ends when the slow roll parameter 
$\epsilon=\frac{1}{2}\left( \frac{V_{\phi}}{V}  \right)^2$ is equal to $1$. In the case of a quadratic potential this means that $\phi_{end}^2=2$. So, given a background
$\phi(t)$, from  $\phi_{end}^2=2$ one calculates $t_{end}$ and thus, all the quantities at that time.

\

Choosing    as in \cite{agullo} the background with  initial condition at the bounce $\phi_B=1.2\times \sqrt{8\pi}\cong 6$, one obtains $N_B\cong 74$. Moreover, for this kind
of potentials inflation ends when $H_{end}\cong 10^{-6}$  \cite{pv}. Then, using the present values of the Hubble parameter and temperature 
$H_0\cong 1.46\times 10^{-42}$ GeV and $T_0\cong 2.34\times 10^{-13}$ GeV one gets
\begin{eqnarray}k_*\cong \frac{80}{g_R^{1/12}T_R^{1/3}}\cong \frac{80}{T_R^{1/3}},
\end{eqnarray}
because $g_R=107$ for $T_R\geq 175$ GeV, $g_R=  90$ for
$200 \mbox{ MeV} \leq  T_R\leq  175 \mbox{ GeV}$, and $g_R= 11$ for $1 \mbox{ MeV} \leq  T_R \leq  200$ MeV \cite{rehagen}.

\

Finally, 
for reheating temperatures consistent with the bounds coming from
nucleosynthesis, i.e., in the range   between  $1$ MeV  and   $10^9$ GeV,   or in our units, for $10^{-21}\leq T_R\leq 10^{-9}$, what constraints the
pivot scale to be in the range
\begin{eqnarray}
8\times 10^4 \leq k_* \leq 8\times 10^8.
\end{eqnarray}

\

 \section{Conclusions}
 We have studied in a simple way, but with great detail, the dynamics of the standar  and  the recent  model of LQC  proposed by Dapor-Liegener model, showing that, contrarily
 to the standard model where the observable modes leave the Hubble radius at very early times,  it is impossible to implement 
 an alternative to the inflationary paradigm as the matter or matter-bounce scenario due to the fact that the universe emerges, in the contracting phase, from a de Sitter regime, meaning that at early times the physical scales,  intead of leaving the Hubble radius,  they enters into it. Therefore, one has to understand the DL model of LQC 
 as an implementation of inflation, which solves the initial singularity problem, but where an slow-roll regime is needed to generate the primordial perturbations.
 \

\section*{Acknowledgments} I would like to thanks Jaume Amor\'os for reading the manuscript and Iv\'an Agull\'o for useful conversations.
This investigation has been supported by MINECO (Spain) grants  MTM2014-52402-C3-1-P and MTM2017-84214-C2-1-P, and  in part by the Catalan Government 2017-SGR-247.


\begin{thebibliography}{}



\bibitem{dl}
A. Dapor and  K. Liegener,
\textit{Cosmological Effective Hamiltonian from full Loop Quantum Gravity Dynamics}, 
(2017) [arXiv:1706.09833].



\bibitem{dl2}
M. Assanioussi, A. Dapor, K. Liegener and T. Pawlowski,
\textit{Emergent de Sitter epoch of the quantum Cosmos}, (1018) 	[arXiv:1801.00768].




\bibitem{ma}
 J. Yang, Y. Ding, and Y. Ma,
\textit{Alternative quantization of the Hamiltonian in loop quantum cosmology II: Including the Lorentz term}, 
 	Phys. Lett. {\bf B682}, 1 (2009) 	[arXiv:0904.4379].





\bibitem{singh} 
A. Ashtekar, T. Pawlowski and P. Singh, 
\textit{Quantum Nature of the Big Bang},
Phys. Rev. Lett. {\bf 96} 141301, (2006)  [arXiv:0602086].
\bibitem{singh1}
A. Ashtekar, T. Pawlowski and P. Singh, 
\textit{Quantum Nature of the Big Bang: Improved dynamics},
Phys. Rev.  {\bf D74}, 084003 (2006) [arXiv:0607039].

\bibitem{singh2}
A. Corichi and P. Singh, 
\textit{Is loop quantization in cosmology unique?}, 
Phys.Rev. {\bf D78} 024034, (2008) [arXiv:0805.0136].
    
    
 \bibitem{singh08}
P. Singh, 
\textit{Transcending Big Bang in Loop Quantum Cosmology: Recent Advances}, 
J. Phys.
Conf. Ser. {\bf 140}, 012005 (2008) [arXiv:0901.1301].



\bibitem{singh09}
P. Singh,
\textit{Are loop quantum cosmos never singular?}, 
Class. Quant. Grav. {\bf 26},  125005 (2009)
[arXiv:0901.2750].   
    



  \bibitem{as}
A. Ashtekar and P. Singh, \textit{Loop Quantum Cosmology: A Status Report},  Class. Quant. Grav. {\bf 28}, 213001 (2011) [arXiv:1108.0893].



\bibitem{dl1}
A. Dapor and  K. Liegener,
\textit{Cosmological Coherent State Expectation Values in LQG I. Isotropic Kinematics}, 
(2017) [arXiv:1710.04015].





\bibitem{singh18}
B-F Li, P. Singh and  A. Wang,
\textit{Towards Cosmological Dynamics from Loop Quantum Gravity},  	Phys. Rev. {\bf D 97}, 084029 (2018)
 	[arXiv:1801.07313].




\bibitem{agullo}
I. Agullo,
\textit{Primordial power spectrum from the Dapor-Liegener model of loop quantum cosmology} (2018)
[arXiv:1805.11356].





\bibitem{we}
E. Wilson-Ewing, \textit{The Matter Bounce Scenario in Loop Quantum Cosmology},  	JCAP {\bf 1303}, 026 (2013)
 	[arXiv:1211.6269].





\bibitem{ha14a} 
 J. Haro and J. Amor\'os,
\textit{Viability of the matter bounce scenario in $F(T)$ gravity and Loop Quantum Cosmology for general potentials}, 
JCAP {\bf 1412},  031 (2014)
[arXiv:1406.0369].  





\bibitem{cw}
 Y.-F. Cai and E. Wilson-Ewing, 
 \textit{Non-singular bounce scenarios in loop quantum cosmology and the effective  field description}, 
 JCAP {\bf 03},  026  (2014) [arXiv:1402.3009]. 
  


\bibitem{wilson0}
E. Wilson-Ewing, \textit{Ekpyrotic loop quantum cosmology} , 
JCAP {\bf 1308}, 015 (2013) 	[arXiv:1306.6582].




\bibitem{agullo1}
I. Agullo, A. Ashtekar and  W. Nelson,
\textit{A Quantum Gravity Extension of the Inflationary Scenario}, 
Phys. Rev. Lett. {\bf 109}, 251301 (2012) 	[arXiv:1209.1609].





\bibitem{meissner}
  K. A. Meissner, 
\textit{Black hole entropy in Loop Quantum Gravity},  
  Class. Quant. Grav.
{\bf 21}, 5245  (2004) [arXiv:0407052].



{{} \bibitem{perez}
A. Ghosh and A. Perez,
{\it Black hole entropy and isolated horizons thermodynamics}, 
PRL {\bf 107}, 241301 (2011) 	[arXiv:1107.1320].


\bibitem{saridakis}
Y-F. Cai, S-H. Chen, J. B. Dent, S. Dutta and E. N. Saridakis,
{\it Matter Bounce Cosmology with the $f(T)$ Gravity},
Class. Quantum Grav. {\bf 28}, 215011 (2011) 	[arXiv:1104.4349].

\bibitem{haro13}
J. Haro,
{\it Cosmological perturbations in teleparallel Loop Quantum Cosmology},
	JCAP{\bf 11}, 068 (2013) 	[arXiv:1309.0352].
}




\bibitem{helling}
R. C. Helling,
{\it Higher curvature counter terms cause the bounce in loop cosmology}, (2009)
 [arXiv:0912.3011].



\bibitem{ds09}
G. Date and S. Sengupta, 
\textit{Effective Actions from Loop Quantum Cosmology: Correspondence with Higher Curvature Gravity},
Class. Quant. Grav. {\bf 26}, 105002 (2009) [arXiv:0811.4023].


\bibitem{ha17}
 J. de Haro and J. Amor\'os,  {\it Bouncing cosmologies via modified gravity in the ADM formalism: Application to Loop Quantum Cosmology},
 (2017) [arXiv:1712.08399]. 




\bibitem{hap}
J. de Haro, L. Arest\'e Sal\'o and S. Pan, \textit{Mimetic Loop Quantum Cosmology}, (2018)
 	[arXiv:1803.09653].

\bibitem{Norbert}
N. Bodendorfer, A. Sch\"afer and J. Schliemann, \textit{On the canonical structure of general relativity with a limiting curvature and its relation to loop quantum gravity}, 
Phys. Rev. {\bf D 97}, 084057 (2018)	
  	[arXiv:1703.10670].


\bibitem{mukhanov1}
A. H. Chamseddine and V. Mukhanov, \textit{Resolving Cosmological Singularities},  JCAP  {\bf 1703},
009 (2017) [arXiv:1612.05860].    
    
    

 
\bibitem{langlois}
   D. Langlois, H. Liu, K. Noui and E. Wilson-Ewing, \textit{Effective loop quantum cosmology as a higher-derivative scalar-tensor theory},
 Class. Quant. Grav. {\bf 34}, 225004 (2017)  [arXiv:1703.10812].  
 



 \bibitem{acs}
 A. Ashtekar, A. Corichi and  P. Singh,    
\textit{       
Robustness of key features of loop quantum cosmology}, 
Phys.Rev. {\bf D 77}, 024046 (2008) 	[arXiv:0710.3565]. 
 
  
\bibitem{singh06}
P. Singh,
\textit{Loop cosmological dynamics and dualities with Randall-Sundrum braneworlds},
Phys.
Rev. {\bf D73}, 063508 (2006) [arXiv:0603043].



\bibitem{svv06}
P. Singh, K. Vandersloot and G. Vereshchagin,
\textit{Non-singular bouncing universes in loopquantum cosmology}, 
Phys. Rev. {\bf D74}, 043510 (2006) [arXiv:0606032].





\bibitem{sst06}
 M. Sami, P. Singh and S. Tsujikawa, 
 \textit{Avoidance of future singularities in loop quantum cosmology}, 
Phys. Rev. {\bf D74}, 043514 (2006) [arXiv:0605113].





\bibitem{bho}
K. Bamba, J. de Haro and  S. D. Odintsov,
\textit{Future singularities and Teleparallelism in Loop Quantum Cosmology}, 
JCAP {\bf 02}, 008 (2013)
[arXiv:1211.2968].


\bibitem{aho}
J. Amor\'os, J. de Haro and S. D. Odintsov,
\textit{
Bouncing Loop Quantum Cosmology from $F(T)$ gravity}, 
Physical Review {\bf D 87}, 104037 (2013)
 	[arXiv:1305.2344].



\bibitem{mielczarek}
  J. Mielczarek, 
\textit{Multi-fluid potential in the loop cosmology},  
  Phys. Lett. {\bf B675}, 273 (2009) [arXiv:0809.2469].


\bibitem{haa}
  J. Haro,  J. Amor\'os and  L. Arest\'e Sal\'o, 
 \textit{ The matter-ekpyrotic bounce scenario in Loop Quantum Cosmology},
   JCAP {\bf 09}, 002 (2017)  	[arXiv:1703.03710].  




\bibitem{aah}
L. Arest\'e Sal\'o, Jaume Amor\'os and J. de Haro
\textit{
Qualitative study in Loop Quantum Cosmology}, 
 	Class. Quant. Grav. {\bf 34}, 235001 (2017) 	[arXiv:1612.05480].


\bibitem{B} 
  R.~H.~Brandenberger,
  \textit{Introduction to Early Universe Cosmology},
  PoS ICFI {\bf 2010}, 001 (2010),
  [arXiv:1103.2271].
  



\bibitem{nb08}
M. Novello and S. E. P. Bergliaffa, \textit{Bouncing Cosmologies},  Phys. Rept. {\bf 463}, 127 (2008) [arXiv:0802.1634]. 


\bibitem{patrick0}
R. Brandenberger and P. Peter, \textit{Bouncing Cosmologies: Progress and Problems  },  
(2016)  	[arXiv:1603.05834].

\bibitem{patrick}
D. Battefeld and P. Peter, \textit{A Critical Review of Classical Bouncing Cosmologies }, 
Phys. Rep. {\bf12},  004 (2014) 	[arXiv:1406.2790].



\bibitem{odintsov0}
S. Nojiri, S.D. Odintsov and V.K. Oikonomou, \textit{Modified Gravity Theories on a Nutshell: Inflation, Bounce and Late-time Evolution  },
 Phys.Rept. {\bf 692}, 1 (2017) [arXiv:1705.11098]. 





\bibitem{bgt}
M. Bojowald, G. Calcagni and  S. Tsujikawa,
\textit{Observational constraints on loop quantum cosmology}
 	Phys. Rev. Lett. {\bf 107}, 211302 (2011)
 	[arXiv:1101.5391].


\bibitem{bgt1}
M. Bojowald, G. Calcagni and  S. Tsujikawa,
\textit{Observational test of inflation in loop quantum cosmology}
JCAP {\bf 11}, 046 (2011) 	[arXiv:1107.1540].


\bibitem{barrau}
A. Barrau,
\textit{Inflation and Loop Quantum Cosmology}, (2011) [arXiv:1011.5516].

\bibitem{sloan}
A. Ashtekar and  D. Sloan, 
\textit{Loop quantum cosmology and slow roll inflation},  	Phys. Lett. {\bf B694}, 108 (2010)
 	[arXiv:0912.4093].




\bibitem{brandenberger}
R. H. Brandenberger
\textit{The Matter Bounce Alternative to Inflationary Cosmology}, (2012) [arXiv:1206.4196]

\bibitem{brandenberger1}
R. H. Brandenberger
\textit{Unconventional Cosmology}, (2012)  	[arXiv:1203.6698].

\bibitem{brandenberger2}
R. H. Brandenberger
\textit{Alternatives to Cosmological Inflation},  (2009) 	[arXiv:0902.4731].




\bibitem{riotto}
A. Riotto, 
\textit{Inflation and the Theory of Cosmological Perturbations}, (2002) 	[arXiv:0210162].



 

\bibitem{lw}
J-L. Lehners and  E. Wilson-Ewing,
\textit{Running of the scalar spectral index in bouncing cosmologies},  	JCAP {\bf 10}, 038 (2015)
 	[arXiv:1507.08112]
 	
 	





\bibitem{wands}
 D. Wands, \textit{Duality Invariance of Cosmological Perturbation Spectra},
 Phys. Rev.
{\bf D 60}, 023507 (1999 ) [arXiv:9809062].






\bibitem{eho}
E. Elizalde, J. Haro and S. D. Odintsov, \textit{Quasi-matter domination parameters in bouncing cosmologies}, 
Phys. Rev. {\bf D 91}, 063522 (2015) 	[arXiv:1411.3475].


\bibitem{agullo2}
I. Agullo, A. Ashtekar and  W. Nelson,
\textit{An Extension of the Quantum Theory of Cosmological Perturbations to the Planck Era}, Phys. Rev. {\bf D87}, 043507 (2013)
 	[arXiv:1211.1354].


\bibitem{agullo3}
I. Agullo, A. Ashtekar and  W. Nelson,
\textit{The pre-inflationary dynamics of loop quantum cosmology: Confronting quantum gravity with observations}
Class. Quant. Grav. {\bf 30}, 085014 (2013) 	[arXiv:1302.0254].






\bibitem{vidotto}
  T. Cailleteau,  A. Barrau, F. Vidotto and J. Grain,
\textit{Consistency of holonomy-corrected scalar, vector and tensor perturbations in Loop Quantum Cosmology}  
   Phys. Rev. {\bf D86}, 087301 (2012)
 [arXiv: 1206.6736].

  
  
  \bibitem{grain}
  T. Cailleteau, J. Mielczarek, A. Barrau and J. Grain, \textit{Anomaly-free scalar perturbations with holonomy corrections in loop quantum cosmology},
  Class. Quant. Grav.  {\bf 29}, 095010 (2012) [arXiv:1111.3535].




\bibitem{lisenfors}
L. Linsefors, T. Cailleteau, A. Barrau and  Julien Grain,
\textit{Primordial tensor power spectrum in holonomy corrected Omega-LQC},
Phys.Rev. {\bf D87},  107503 (2013) 	[arXiv:1212.2852].



\bibitem{grain1}
A. Barrau, M. Bojowald, G. Calcagni, J. Grain and M. Kagan,
\textit{Anomaly-free cosmological perturbations in effective canonical quantum gravity}, 
 	JCAP {\bf 05}, 051 (2015) 	[arXiv:1404.1018].






{
\bibitem{bojowald0}
M. Bojowald, S. Brahma and  J. D. Reyes,
\textit{Covariance in models of loop quantum gravity: Spherical symmetry}, 
Phys. Rev. {\bf D 92}, 045043 (2015)	[arXiv:1507.00329]}

\bibitem{bojowald1}
M. Bojowald, S. Brahma, U. Buyukcam and  F. D'Ambrosio, 
\textit{Hypersurface-deformation algebroids and effective space-time models}, Phys. Rev. {\bf D 94}, 104032 (2016)
	[arXiv:1610.08355].
	
\bibitem{bojowald2}
M. Bojowald, S. Brahma, and D. Yeom, 
\textit{ Effective line elements and black-hole models in canonical (loop) quantum gravity}, 
(2018)	[arXiv:1803.01119].





{{}
\bibitem{hae18}
J. de Haro, L. Arest\'e Sal\'o and  E. Elizalde,
{\it Cosmological perturbations in a class of fully covariant modified theories: Application to models with the same background as standard LQC},
EPJC {\bf 78}, 712 (2018) 	[arXiv:1806.07196]. }


\bibitem{basset}
  B.A. Bassett, S. Tsujikawa and D. Wands, 
  \textit{Inflation Dynamics and Reheating}, 
  Rev. Mod. Phys. {\bf 78} , 537 (2006) [arXiv:0507632].

\bibitem{planck}
  P.A.R. Ade
et al.
[Planck Collaboration], 
\textit{Planck 2013 results. XXII. Constraints on inflation},
Astron. Astrophys. {\bf 571}, A22 (2014) [arXiv:1303.5082].

\bibitem{planck1}
   P.A.R. Ade
et al.,
\textit{ A Joint Analysis of BICEP2/Keck Array and Planck Data},
Phys. Rev. Lett.
{\bf 114}, 101301 (2015) [arXiv:1502.00612].
  










\bibitem{rehagen}
T. Rehagen and G. B.  Gelmini, 
\textit{Low reheating temperatures in monomial and binomial inflationary potentials},
JCAP {\bf  06}, 039 (2015)	 [arXiv:1504.03768].



\bibitem{linde}
L. Kofman, A. Linde  and A. Starobinsky, \textit{Towards the Theory of Reheating After Inflation},
Phys. Rev. {\bf D56}, 3258  (1997)	[arXiv:9704452].
\bibitem{turner}
M.S. Turner,   
\textit{Coherent scalar-field oscillations in an expanding universe},
Phys. Rev. {\bf D 28}, 1243 (1983).
\bibitem{ford}
L. H.  Ford,  
\textit{Gravitational particle creation and inflation},  Phys. Rev. {\bf D 35}, 2955 (1987).

\bibitem{pv}
P. J. E. Peebles and A. Vilenkin,
\textit{ Quintessential inflation}, 
 	Phys. Rev. {\bf D 59},  063505 (1999)
		[arXiv:9810509].	




\end{thebibliography}
\end{document}